\documentclass[a4paper,12pt]{article}
\usepackage[margin=2cm]{geometry}
\usepackage[utf8]{inputenc}
\usepackage[T2A]{fontenc}
\usepackage[english]{babel}
\usepackage{amsmath}
\usepackage{amssymb}
\usepackage{bm}
\usepackage[colorlinks=true,linkcolor={blue},citecolor={green},urlcolor={red}]{hyperref}
\usepackage[dvipdf]{graphicx}
\usepackage{authblk}
\usepackage{indentfirst}
\usepackage{cite}

\usepackage{bibmacros_shortened_utf8}

\newcommand{\Dfrac}[2]{\frac{d#1}{d#2}}

\newcommand{\DPfrac}[2]{\frac{\partial#1}{\partial#2}}

\newcommand{\TDfrac}[3]{\left(\frac{\partial#1}{\partial#2}\right)_{\!\!#3}}

\begin{document}

\title{Excitation of turbulence in accretion disks of binary stars by non-linear perturbations}

\author[1]{E.~P.~Kurbatov\thanks{kurbatov@inasan.ru}}
\author[1]{D.~V.~Bisikalo\thanks{bisikalo@inasan.ru}}
\affil[1]{Institute of Astronomy of Russian Academy of Sciences}

\maketitle

\begin{abstract}

Accretion disks in binary systems can experience hydrodynamic impact at inner as well as outer edges. The first case is typical for protoplanetary disks around young T Tau stars. The second one is typical for circumstellar disks in close binaries. As a result of such an impact, perturbations with different scales and amplitudes are excited in the disk. We investigated the nonlinear evolution of perturbations of a finite, but small amplitude, at the background of sub-Keplerian flow. Nonlinear effects at the front of perturbations lead to the formation of a shock wave, namely the discontinuity of the density and radial velocity. At this, the tangential flow in the neighborhood of the shock becomes equivalent to the flow in in the boundary layer. Instability of the tangential flow further leads to turbulization of the disk. Characteristics of the turbulence depend on perturbation parameters, but $\alpha$-parameter of Shakura-Sunyaev does not exceed $\sim 0.1$.

\bigskip\noindent
PACS numbers:\,
47.27.-i, 
97.10.Gz, 
97.80.-d 
\end{abstract}

\section{Introduction}

Turbulent viscosity is the most effective mechanism of the angular momentum transfer in astrophysical disks \cite{Shakura1972AZh....49..921S,Shakura1973A&A....24..337S,Lynden-Bell1974MNRAS.168..603L}. The turbulence is caused by development of instability of small perturbations. In accretion disks of nonmagnetic stars, as well as in the protoplanetary disks (in cold non-ionized gas), the turbulence can develop from instabilities of the hydrodynamic type only.

The stability of linear hydrodynamic perturbations in Keplerian disks is studied in numerous papers (see, for instance, references in \cite{Godon1999ApJ...521..319G}). Both radial and non-axisymmetric perturbations of the modal type turn out to be stable according to the Rayleigh criterion \cite{Rayleigh1917RSPSA..93..148R,Landau1987theorphys-6}. The reason for this is spectral stability of the Keplerian flow. There are non-modal type perturbations or transient ones, which display a growth in the linear stage (see a review in \cite{Razdoburdin2015PhyU...58.1031R}). However, the problem of the mechanism of their continuous generation in the Keplerian flow, as well as the problem of efficiency of removal of angular momentum, on large time scales needs further study.

There are at least two classes of astrophysical disks in which accretion flow experiences an external action: the disks around components of close binary stars and disks around young stellar systems (protoplanetary ones). In the objects of the first kind, the stream from the donor-star acts upon the outer part of the accretion disk, generating a tangential discontinuity and associated with the latter shock wave, the so-called \glqq{hot line}\grqq\ \cite{Bisikalo1998MNRAS.300...39B,Boyarchuk2002mtcb.book.....B}. In the objects of the second kind act detached shocks from the components of the binary system \cite{Kaygorodov2010} or shockless perturbations, if the star moves with a subsonic velocity \cite{Sytov2016AstRep}. The last affect the protoplanetary disk at the inner side. As a result, nonlinear waves of different amplitude and scale may be exited in the accretion disk. Although in a certain conditions the nonlinear waves may provide a conducive background for linear instability and the turbulence formation \cite{Kurbatov2014}, it is not the case in the protoplanetary disks, since the nonlinear perturbations are too weak.

In hydrodynamics, the mechanism of formation of discontinuities in nonlinear waves is known \cite{Zeldovich1966shockwaves.book,Landau1987theorphys-6}. This phenomenon is due to the “breaking” of the wave profile. If the discontinuity is a shock wave, dissipative effects will lead to the damping of the perturbation with time. On the other hand, in the accretion flows the tangential velocity of a gas can many times larger than the sound velocity. Therefore, it can not be ruled out that, for certain perturbation parameters the rollover of the profile may cause a tangential discontinuity. In this case, the difference of energy at different sides of the discontinuity will transform into kinetic energy of the turbulence.

In the present paper, we investigate whether evolution of nonlinear waves in the accretion disk may lead to the appearance of unstable configurations and, as a result, to turbulence. we We perform an analysis of nonlinear perturbations in sub-Keplerian disks with inhomogeneous distribution of density and pressure. The paper is organized as follows: in the Section 2 we describe the structure of accretion flow in semi-detached and young binary systems. In Section 3 we present the model of an equilibrium sub-Keplerian disk and a briefly analyze linear perturbations in the latter. In the fourth Section we explore evolution of nonlinear radial perturbations in accretion disks. In Section 5 we apply our results to two simple models of accretion disks. Conclusions and deductions are presented in the last Section.

\section{The structure of accretion flows in young binaries with circumstellar envelope and in close binaries}

Accretion disks in binary systems have the distribution of angular momentum close to the Keplerian one. Velocity of tangential motion in them can be hundreds times larger than the sound velocity (in the central regions). Often, accretion disks experience hydrodynamic action from accretion flows, encounter with the streams of incoming gas or detached shock waves. Such effects, along to large tangential velocities, can be potential sources of turbulence. Below, we will consider the flow patterns for two classes of typical binary stars: young stars surrounded by a protoplanetary disk, and close binary systems.

The envelope of a T Tauri binary star is a remnant of the protostellar cloud from which the matter falls and forms a Keplerian protoplanetary disk. In the central part of the disk around the binary forms a gas-free region, a “gap”. The reason for the existence of the gap can be both tidal action (Lindblad resonances) of the binary \cite{Artymowicz1994ApJ...421..651A} and the effects of the detached shock waves produced by the components of the system \cite{Kaygorodov2010}. Typical gap width is of the order of several orbital separations (see, e.g., \cite{Jensen1997AJ....114..301J,Najita2003ApJ...589..931N}). Inside the gap, around each of the stars forms its own circumstellar accretion disk and in the case of the supersonic orbital motion of components (which is typical of such systems), detached shock waves appear in the front of the stars \cite{Fateeva2011Ap&SS.335..125F,Sytov2011AstRep,GomezDeCastro2013ApJ...766...62G}. In the region of the inner disk boundary, cold and dense gas of the Keplerian flow is contiguous to the hot and rarefied gas of the gap. Keplerian velocity in this region can exceed the sound velocity by many dozen times. At the same time, the impact of the detached shock wave upon the disk is concentrated on a small part of the boundary. Since the impact has a shock nature, and it is quite intense, it is obvious that it will result in destruction of the Keplerian flow in the vicinity of the contact. Blurring of this region by differential rotation of the gas can lead to the formation of a turbulent layer over a time of the order of one revolution of the disk at the radius of the inner boundary.

In close binary systems, one of the components (the donor) fills its Roche lobe. This results in the flow of the matter onto the second component (accretor). Under the influence of the dissipative processes, a disk forms around the accretor. The former has a Keplerian angular momentum distribution. Typical is the case when the accretor is more massive and is a $\sim 1~M_\odot$ white dwarf, while the mass of the donor is lower. The radius of the accretion disk comprises the tenth of the orbital separation, while velocity of Keplerian motion even at the outer edge of the disk can exceed the sound speed be several tens of times. The region of the outer edge of the disk where the stream from the donor interacts with accretion disk, is a tangential discontinuity \cite{Bisikalo1998MNRAS.300...39B,Boyarchuk2002mtcb.book.....B,Bisikalo2003AstRep-hotline}. The step of tangential velocity at the discontinuity can be of the order of the Keplerian velocity at the edge of the disk. It is known that such discontinuities are unstable for many types of perturbations \cite{Fridman2008PhyU...51..213F}. Because of the high flow velocities, the instability must have a drift character and manifest itself as a turbulent wake from the tangential discontinuity in the direction of rotation of the gas.

In both examples considered above, the turbulence is excited in the Keplerian flow: in the inner region of the protoplanetary disks and in the outer region of the circumstellar disk in semi-detached binary systems. The turbulence, however, engulfs only a small part of the disk flow. The problem of possible further propagation of turbulence into entire volume of the accretion disk is of particular interest, since if this is a case, existing nonlinear perturbations can be considered as the source of turbulence of nonmagnetic accretion disks.

The effect of detached shock waves upon internal boundary of the protoplanetary disk should not only lead to the destruction of the boundary Keplerian flow, but also to the propagation of the perturbations with different amplitudes and scales inside the disk. A similar situation should take place also in the disks of close binaries --- slight perturbations of the rate of mass loss by the donor star or characteristics of the flow at the edge of the accretion disk (for example, due to the intrinsic precession of the accretion disk) can cause changes in the position and intensity of the tangential discontinuity, which, in turn, will disturb the accretion flow.

In the Introduction, we mentioned that small amplitude perturbations are either stable (for instant, radial ones are stable by the Rayleigh criterion \cite{Balbus1996ApJ...467...76B}), or display only a weak instability (some nonaxisymmetric perturbations are unstable \cite{Razdoburdin2015PhyU...58.1031R}, but they have rather large time scale of growth). On the contrary, non-linear waves can participate in the universal scenario of generation of shock waves and tangential discontinuities which is realized in any environment with ordinary equation of state \cite{Zeldovich1966shockwaves.book,Landau1987theorphys-6}. At the conditions of fast shear flow these discontinuities will be unstable and will decay directly, generating turbulence.

A nonlinear wave is a propagating perturbation of density, pressure, and velocity, with a finite amplitude. Two factors are responsible for the formation of a discontinuity in a nonlinear wave: (i) in the areas of enhanced pressure, propagation velocity of perturbations is higher than in the low-pressure regions; (ii) velocity of perturbations is added with the velocity of the background flow. So, if consider the leading edge of a nonlinear wave, where the pressure and velocity vary from the maximum to the minimum values, the region of the maximum will catch up with the minimum region. This will lead to the steepening of the wave profile and to its “rollover” and formation of a discontinuous solution in a finite time. This pattern becomes somewhat more complicated if one takes into account the differential rotation of the disk matter. For instance, the wave solution for small radial perturbations exists only in the region of the Keplerian disk, where the frequency of the wave is not less than the angular velocity of the matter of the disk \cite{Goldreich1979ApJ...233..857G}. However, we will show below that the scenario of formation of discontinuities can take place in the rotating flows too.

Next, we will perform a brief analysis of the linear perturbations in the accretion disks and will consider the mechanism of formation of discontinuities in the nonlinear perturbations in such disks.

\section{Linear perturbations in the accretion disk}

\subsection{Equilibrium disk}
\label{sec:equillibrium}

Let consider a stationary disk flow, symmetric respective to the $r = 0$ axis and $z = 0$ plane. The equation of radial balance of forces in the mid-plane of the disk:
\begin{equation}
  \frac{1}{\rho_0}\,\DPfrac{p_0}{r}
  = - r \Omega_\mathrm{K}^2 + \frac{\lambda_0^2}{r}  \;,
\end{equation}
where $\rho_0$, $p_0$, $\lambda_0$ are, respectively, density, pressure, and angular momentum in the  $z = 0$ plane; $\Omega_\mathrm{K}$ is Keplerian angular velocity
\begin{equation}
  \Omega_\mathrm{K}
  = \Omega_{\mathrm{K}\ast} \left( \frac{r}{r_\ast} \right)^{-3/2}  \;.
\end{equation}
By \glqq{$\ast$}\grqq\ we will mark the values of variables in the point $r_\ast$ at $z = 0$. In particular, $\Omega_{\mathrm{K}\ast} = (G M / r_\ast^3)^{1/2}$. The gas pressure usually decreases with the radius. Therefore the centrifugal acceleration is not exactly equal to the gravitational one, but is somewhat lower \cite{Pringle1981ARA&A..19..137P}. Let the disk to be sub-Keplerian, if the balance of gravitational and centrifugal force is violated by a small amount \cite{Armitage2007astro.ph..1485A}:
\begin{equation}
  \label{eq:subkeplerian_disk}
  r \Omega_\mathrm{K}^2 - \frac{\lambda_0^2}{r^3}
  = \chi\,\frac{H^2}{r^2}\,r \Omega_\mathrm{K}^2  \;,
\end{equation}
where $H$ is semi-thickness of the disk; $\chi$ --- a dimansionless parameter. Let distribution of the
matter to be isentropic with the exponent $\gamma$, distribution of density $\rho_0$ on radius is a power law with exponent $d$. Then, for the pressure, density and sound velocity one may write down
\begin{gather}
  \rho_0(r)
  = \rho_\ast \left( \frac{r}{r_\ast} \right)^{-d}  \;,  \\
  p_0
  = p_\ast \left( \frac{\rho_0}{\rho_\ast} \right)^\gamma
  = p_\ast \left( \frac{r}{r_\ast} \right)^{-\gamma d}  \;,  \\
  c_0^2(r)
  = c_\ast^2 \left( \frac{r}{r_\ast} \right)^{-(\gamma-1)d}  \;,
\end{gather}
where $c_\ast^2 = \gamma p_\ast / \rho_\ast$. Then the equation of the radial balance of forces becomes
\begin{equation}
  \label{eq:radial_balance}
  d c_\ast^2 \left( \frac{r}{r_\ast} \right)^{-(\gamma-1)d+3}
  = \chi \Omega_{\mathrm{K}\ast}^2 H^2(r)  \;.
\end{equation}

Let assume that the vertical structure of the disk is isothermal. In a thin disk, $H \ll r$, the density is distributed as
\begin{equation}
  \rho(r, z)
  = \rho_0(r)\,\operatorname{exp}\left[ - \frac{z^2}{2 H^2(r)} \right]  \;,
\end{equation}
where semi-thickness of the disk is determined from the condition of the vertical balance of forces
\begin{equation}
  \label{eq:vertical_balance}
  H^2(r)
  = \frac{c_0^2(r)}{\Omega_\mathrm{K}^2(r)}
  = \frac{c_\ast^2}{\Omega_{\mathrm{K}\ast}^2} \left( \frac{r}{r_\ast} \right)^{-(\gamma-1)d+3}  \;.
\end{equation}
From this expression, using Eq. (\ref{eq:radial_balance}) we may derive the factor $\chi$ in the Eq. (\ref{eq:subkeplerian_disk}):
\begin{equation}
  \chi = d  \;.
\end{equation}
For the accepted distribution $\rho(r, z)$ the dependence of the surface density on the radius will be a power law:
\begin{equation}
  \Sigma(r)
  = 2 \int_0^\infty dz\,\rho(r, z)
  = \sqrt{2 \pi}\,\frac{\rho_\ast c_\ast}{\Omega_{\mathrm{K}\ast}} \left( \frac{r}{r_\ast} \right)^{[-(\gamma+1)d+3]/2}  \;.
\end{equation}

In the realistic models of the accretion disks, surface density declines with radius, approximately, as $1/r^{0 \dots 1}$ (see \cite{Armitage2007astro.ph..1485A}  and references therein). For instance, in the Shakura-Sunyaev disk model $\Sigma \propto r^{-3/5}$ \cite{Shakura1973A&A....24..337S}. In the paper \cite{Armitage2015arXiv150906382A}  as a typical dependence is assumed $\sigma \propto r^{-1}$. In our model, we shall assume $\Sigma \propto r^{-1/2}$. Below, this model will allow to obtain more simple expressions for some of dependences, while remaining qualitatively consistent with more complete models.

The slope of the surface density distribution in combination with the value of the adiabatic exponent $\gamma$, determines the values of exponents in power-laws for remaining thermodynamical quantities: volumetric density, pressure, and sound velocity. Further in the present study, we will consider the gas as monatomic. Thus, we obtain
\begin{equation}
  \gamma = 5/3  \;,\qquad
  d = 3/2  \;,\qquad
  (\gamma-1)d = 1  \;.
\end{equation}

Let write down final expressions for equilibrium distributions:
\begin{gather}
  \rho_0
  = \rho_\ast \left( \frac{r}{r_\ast} \right)^{-3/2}  \;,  \\
  c_0
  = c_\ast \left( \frac{r}{r_\ast} \right)^{-1/2}  \;,  \\
  \label{eq:subkeplerian_momentum}
  \lambda_0
  = \left( 1 - \frac{3}{2}\,\frac{c_\ast^2}{r_\ast^2 \Omega_{\mathrm{K}\ast}^2} \right)^{1/2}
      \left( \frac{r}{r_\ast} \right)^{1/2} r_\ast^2 \Omega_{\mathrm{K}\ast}  \;,  \\
  H
  = \frac{c_\ast}{\Omega_{\mathrm{K}\ast}}\,\frac{r}{r_\ast}  \;.
\end{gather}
The typical values of the gas velocity in the disks, according to numerical models, are equal to tens of sound velocities \cite{Bisikalo1998MNRAS.300...39B}. Thus, correction to the Keplerian momentum in expression (\ref{eq:subkeplerian_momentum}) does not
exceed one per cent.

\subsection{Linear perturbations}
\label{sec:linear_pert}

Let consider the dynamics of gas flow in the mid-plane of the disk and confine ourselves to perturbations of the equilibrium configurations that (i) are symmetric about the axis $r = 0$; (ii) do not depend on the vertical coordinate z; (iii) have a zero velocity in the direction of the z-axis. The system of equations that describes such a class of flows has the form:
\begin{gather}
  \label{eq:continuity_general}
  \DPfrac{\rho}{t}
    + \frac{1}{r}\,\DPfrac{(r \rho v)}{r}
  = 0  \;,  \\
  \label{eq:radial_momentum_general}
  \DPfrac{v}{t}
    + v\,\DPfrac{v}{r}
  = - \frac{1}{\rho}\,\DPfrac{p}{r}
    - r \Omega_\mathrm{K}^2
    + \frac{\lambda^2}{r^3}  \;,  \\
  \label{eq:angular_momentum_general}
  \DPfrac{\lambda}{t}
    + v\,\DPfrac{\lambda}{r}
  = 0  \;,
\end{gather}
where $v$ is radial velocity; $\lambda$ --- angular momentum per unit mass.

Let assume that the system of equations (\ref{eq:continuity_general})--(\ref{eq:angular_momentum_general}) describes a small perturbation of the equilibrium configuration, which was determined in the previous section. The perturbation itself we describe as a radial displacement of gas elements. Let denote by $\xi(t, r)$ the value of radial displacement of the particle whose position at the time $t$ is $r$, i.e., the radial coordinate of the nonmoved particle is $r - \xi(t, r)$. From the conservation of the angular momentum law in form (\ref{eq:angular_momentum_general}) it follows that the moment of each element of the gas is conserved, i. e., we may write down
\begin{equation}
  \label{eq:angular_momentum_expansion}
  \lambda^2(t, r)
  = \lambda_0^2\bigl(r - \xi(t, r)\bigr)
  = \lambda_0^2(r) - \Dfrac{\lambda_0^2(r)}{r}\,\xi(t, r) + \mathcal{O}\!\left( r \Omega_\mathrm{K}^2 \frac{\xi^2}{r^2} \right)  \;.
\end{equation}
Here we took into account that $|\xi| \ll r$. In our model the derivative $d\lambda_0^2/dr$ does not depend on the coordinate and is equal to
\begin{equation}
  \Dfrac{\lambda_0^2}{r}
  = \left( 1 - \frac{3}{2}\,\frac{c_\ast^2}{r_\ast^2 \Omega_{\mathrm{K}\ast}^2} \right) r_\ast^3 \Omega_{\mathrm{K}\ast}^2  \;.
\end{equation}

Let require that perturbations of interest have a small spatial scale of variability. Then we can neglect the \glqq{geometric}\grqq\ term $\rho v/r$ in the equation of continuity, which is determined by the cylindrical geometry of the problem, and the advective term in (\ref{eq:continuity_general}) will take a Cartesian form. This approximation was used by many authors in the analysis of perturbations in the accretion disks (see, for instance, \cite{Lubow1993ApJ...409..360L,Balbus1996ApJ...467...76B}).

Let denote perturbed values by an index \glqq{$1$}\grqq. Then, after linearization of equation of continuity (\ref{eq:continuity_general}) over small perturbations, we get
\begin{equation}
  \label{eq:continuity_pert}
  \DPfrac{\rho_1}{t}
    + \DPfrac{(\rho_0 v_1)}{r}
  = 0  \;.
\end{equation}
By means of (\ref{eq:angular_momentum_expansion}) linearization of the Euler equation (\ref{eq:radial_momentum_general}) will be
\begin{equation}
  \label{eq:radial_momentum_pert}
  \DPfrac{v_1}{t}
    + \frac{1}{\rho_0}\,\DPfrac{p_1}{r}
    - \Dfrac{p_0}{r}\,\frac{\rho_1}{\rho_0^2}
    + \frac{1}{r^3 }\,\Dfrac{\lambda_0^2}{r}\,\xi
  = 0  \;.
\end{equation}
Angular momentum conservation law was already taken into account explicitly when we employed
expansion (\ref{eq:angular_momentum_expansion}). The gradients of thermodynamical values for adiabatic equation of state have the
form:
\begin{gather}
  \Dfrac{p_0}{r}
  = c_0^2\,\Dfrac{\rho_0}{r}  \;, \\
  p_1
  = c_0^2 \rho_1  \;,  \\
  \Dfrac{c_0^2}{r}
  = (\gamma - 1)\,\frac{c_0^2}{\rho_0}\,\Dfrac{\rho_0}{r}  \;.
\end{gather}

To close the system (\ref{eq:continuity_pert}), (\ref{eq:radial_momentum_pert}), let use the relation between dislocation, density and velocity \cite{Lynden-Bell1974MNRAS.168..603L}:
\begin{gather}
  \label{eq:rho_xi}
  \rho_1
  = - \DPfrac{(\rho_0 \xi)}{r}  \;,  \\
  \label{eq:v_xi}
  v_1
  = \DPfrac{\xi}{t}  \;.
\end{gather}
It easy to show that, if these definitions are used, continuity equation (\ref{eq:continuity_pert})  is satisfied automatically. Inserting (\ref{eq:rho_xi}) and (\ref{eq:v_xi})  into the radial component of equations of motion equation (\ref{eq:radial_momentum_pert}) we get a closed equation for the displacement:
\begin{equation}
  \label{eq:xi}
  \DPfrac{^2 \xi}{t^2}
  = \frac{c_0^2}{\rho_0}\,\DPfrac{^2 (\rho_0 \xi)}{r^2}
    + (\gamma - 2)\,\frac{c_0^2}{\rho_0^2}\,\DPfrac{\rho_0}{r}\,\DPfrac{(\rho_0 \xi)}{r}
    - \frac{1}{r^3}\,\Dfrac{\lambda_0^2}{r}\,\xi  \;.
\end{equation}
The algorithm of solution of this equation is presented in the Appendix A. Solution has the form
of a progressing wave:

\begin{equation}
  \label{eq:xi_solution}
  \xi
  = r_\ast \Re\!\left\{ C e^{\mp i \omega t} \left( \frac{r}{r_\ast} \right)^{7/4}
    H_\nu\!\left[ \frac{2 r_\ast \omega}{3 c_\ast} \left( \frac{r}{r_\ast} \right)^{3/2} \right] \right\}  \;,
\end{equation}
where $H_\nu$ is the Hankel function of the first kind of the order $\nu$; $C$ is a complex constant; $\Re\{\cdot\}$ --- real part of its argument. The order is defined as
\begin{equation}
  \nu
  = \frac{1}{6} \left( 16\,\frac{r_\ast^2 \Omega_{\mathrm{K}\ast}^2}{c_\ast^2} - 23 \right)^{1/2}  \;.
\end{equation}
It is necessary to note that the solution of Eq. (\ref{eq:xi_solution}) has a wave form only in the case, when the following inequality keeps (see Appendix A):
\begin{equation}
  \label{eq:wave_constraint}
  \omega^2
  > \left( 1 - \frac{3}{2}\,\frac{c_\ast^2}{r_\ast^2 \Omega_{\mathrm{K}\ast}^2} \right) \Omega_\mathrm{K}^2
  \approx \Omega_\mathrm{K}^2  \;.
\end{equation}
A similar result was obtained in \cite{Goldreich1979ApJ...233..857G} by WKB-method. The last inequality may be rewritten as
\begin{equation}
  \frac{r}{r_\ast}
  \gtrsim \left( \frac{\Omega_{\mathrm{K}\ast}}{\omega} \right)^{2/3}  \;.
\end{equation}
Thus, perturbation wave with a given frequency $\omega$ may progress in the outer part of the disk only, where $\omega > \Omega_\mathrm{K}$.

At large distance from the center dependence of radial displacement on the radius gets a simpler form (see Appendix A):
\begin{equation}
  \label{eq:xi_approx_solution}
  \xi
  \approx |C|\,r_\ast \left( \frac{3 c_\ast}{\pi r_\ast \omega} \right)^{1/2}
      \frac{r}{r_\ast}
      \,\cos\!\left[ \frac{2 r_\ast \omega}{3 c_\ast} \left( \frac{r}{r_\ast} \right)^{3/2} \mp \omega t + (\dots) \right]  \;,
\end{equation}
where $(\dots)$  is an insignificant phase addition. Using (\ref{eq:rho_xi}) and (\ref{eq:v_xi}) we can obtain expressions for perturbations of density and velocity:
\begin{gather}
  \label{eq:rho_approx}
  \rho_1
  \approx |C|\,\rho_\ast \left( \frac{3 r_\ast \omega}{\pi c_\ast} \right)^{1/2}
      \,\sin\!\left[ \frac{2 r_\ast \omega}{3 c_\ast} \left( \frac{r}{r_\ast} \right)^{3/2} \mp \omega t + (\dots) \right]  \;,  \\
  \label{eq:v_approx}
  v_1
  \approx \pm |C|\,c_\ast \left( \frac{3 r_\ast \omega}{\pi c_\ast} \right)^{1/2}
      \frac{r}{r_\ast}
      \,\sin\!\left[ \frac{2 r_\ast \omega}{3 c_\ast} \left( \frac{r}{r_\ast} \right)^{3/2} \mp \omega t + (\dots) \right]  \;.
\end{gather}

Let get an expression for perturbation of the angular momentum. Using expansion (\ref{eq:angular_momentum_expansion}) one may write down
\begin{equation}
  \lambda_1
  \approx - \Dfrac{\lambda_0^2}{r}\,\frac{\xi}{2 \lambda_0}  \;.
\end{equation}
Applying approximate relation (\ref{eq:xi_approx_solution}), we get
\begin{equation}
  \label{eq:lambda_approx}
  \lambda_1
  \approx - \frac{|C|}{2}\,r_\ast^2 \Omega_{\mathrm{K}\ast}
    \left( 1 - \frac{3}{2}\,\frac{c_\ast^2}{r_\ast^2 \Omega_{\mathrm{K}\ast}^2} \right)^{1/2}
    \left( \frac{3 c_\ast}{\pi r_\ast \omega} \right)^{1/2}
      \left( \frac{r}{r_\ast} \right)^{1/2}
      \,\cos\!\left[ \frac{2 r_\ast \omega}{3 c_\ast} \left( \frac{r}{r_\ast} \right)^{3/2} \mp \omega t + (\dots) \right]  \;.
\end{equation}

It is easy to estimate the typical space scale of the wave $\ell$ from oscillating part of solution (\ref{eq:xi_approx_solution}). As the definition of $\ell$ let write down the equation for the phase:
\begin{equation}
  \label{eq:phase}
  0
  = \left( \DPfrac{}{t} \pm c_0\,\DPfrac{}{r} \right) \cos[ \dots ]
  = \pm \left( \omega - \frac{c_0}{\ell} \right) \sin[ \dots ]  \;,
\end{equation}
where $[\dots]$ stands for the term in square brackets in (\ref{eq:xi_approx_solution}). Then
\begin{equation}
  \label{eq:wavelength}
  \ell
  = \frac{c_0}{\omega}
  = \frac{\Omega_\mathrm{K}}{\omega}\,H  \;.
\end{equation}
Comparing this expression with condition (\ref{eq:wave_constraint}) one may note that the perturbations have a wave form, if their \glqq{wavelength}\grqq\ $\ell$ does not exceed semi-thickness of the disk. In other words, the waves can not propagate into the region $\ell > H$.

Linear perturbations in the rotating flows were considered by many authors. For instance, in the classical paper \cite{Goldreich1979ApJ...233..857G} by WKB-method, was obtained a dispersion relation for the waves:
\begin{equation}
  \omega^2
  = \Omega_\mathrm{K}^2 + c_0^2 k^2  \;,
\end{equation}
where $k$ is the parameter of the WKB-method, which is interpreted as the wave number. Such an interpretation implies that the sense of the value $k^{-1}$  is the length of the perturbation wave. This value is related to $\ell$ from (\ref{eq:wavelength}) as
\begin{equation}
  \label{eq:wavelength_wkb}
  k^{-1}
  = \frac{\ell}{(1 - \Omega_\mathrm{K}^2/\omega^2)^{1/2}}  \;.
\end{equation}
In the depth of the wave zone ($\omega \gg \Omega_\mathrm{K}$) the values of both expressions comply with precision up to value of the order $\Omega_\mathrm{K}^2/\omega^2$. However, in the region $\omega \approx \Omega_\mathrm{K}$ definition of wave length by expression (\ref{eq:wavelength_wkb})  provides a physically unacceptable result, since $k^{-1} \to \infty$. Below, as the length of perturbation wave we will bear in mind expression (\ref{eq:wavelength}), which, in essence, is an estimate of the distance between neighboring maxima of the wave.

Description of the dynamics of small perturbations in the terms of radial displacement $\xi$ is completely equivalent to the usual approach to the analysis of linear waves in the accretion disks, examples of which can be found in \cite{Lubow1993ApJ...409..360L,Zhuravlev2007AstL...33..740Z} and many other papers. This method was used repeatedly earlier, see, e.g., description of the general approach to the analysis of Lagrangian perturbations in \cite{Lynden-Bell1974MNRAS.168..603L}. In the present study, the use of the Lagrangian approach is preferable, since it allows to express more easily the initial conditions for the analysis of nonlinear perturbations, which is carried out in the next Section.

\section{Formation of discontinuities}

It is known that nonlinear perturbations propagating in the gas can evolve with time into shock waves \cite{Landau1987theorphys-6,Zeldovich1966shockwaves.book}. Usually, the proof of this statement is based on the solution in the form of a simple Riemann wave for one-dimensional gas dynamics \cite{Landau1987theorphys-6}. In the polytropic flow, described by a simple wave, all gasdynamic quantities are expressed through one variable. As it can be seen from the expressions (\ref{eq:rho_approx}) and (\ref{eq:v_approx}), in the case of a perturbation of inhomogeneous medium, density and velocity distributions depend on the coordinate differently and, therefore, they do not form a simple wave. However, it is possible to show that in this case the perturbation can also evolve into a shock wave. If there is a nonzero tangential velocity, on both sides of the shock wave forms a flow, equivalent to the flow in the boundary layer. To prove this assertion, we use the formulation of equations for one-dimensional gas flow presented in the book \cite{Landau1987theorphys-6}.

First, we note that we are interested in perturbations from the wave part of the accretion disk, with wavelength which does not exceed the semi-thickness of the disk (see the previous Section). Let us consider, for definitiveness, a perturbation in the form of a wave (\ref{eq:rho_approx}), (\ref{eq:v_approx}). If nonlinear effects would be taken into account, the form of this perturbation will be distorted, but such characteristics as the amplitude and typical scale of variability (\glqq{wavelength}\grqq\ $\ell$) will be about the same as in the linear case, at least up to the beginning of the formation of discontinuities.

Let estimate the contribution of various terms to the linearized Euler equation (\ref{eq:radial_momentum_pert}). The amplitude of velocity perturbations we normalize by the Mach number $\mathcal{M}$: $|v_1| \sim \mathcal{M} c_0$. Then the amplitude of density perturbations can be estimated as $|\rho_1| \sim \mathcal{M} \rho_0$. During one wave period, every gas element oscillates along the direction of propagation of the perturbation with an amplitude of the order of $|v_1| \ell/c$. This value can be used as an estimate of the displacement amplitude $|\xi|$. It follows from these estimates that in the progressing wave, the ratio of pressure force and the sum of the gravitational and centrifugal forces is
\begin{equation}
  \frac{c_0^2}{\rho_0}\,\frac{|\rho_1|}{\ell}
  : \Omega_\mathrm{K}^2 |\xi|
  \;\;\sim\;\;
  \frac{\mathcal{M} c_0^2}{\ell}
  : \Omega_\mathrm{K}^2 \mathcal{M} \ell
  \;\;\sim\;\;
  1
  : \frac{\ell^2}{H^2}  \;.
\end{equation}
In the wave part of the disk $\ell < H$,  therefore, in order to avoid complicating of further presentation, we neglect the influence of gravitational and centrifugal acceleration upon dynamics of the wave.

It may seem that, at the background of our approximations, all effects caused by differential rotation disappear. Indeed, we demanded that perturbations have sufficiently small scale of variability in comparison with the typical disk radius and the perturbation amplitude is small enough that it would be possible to interpret the sum of the gravitational and centrifugal accelerations in terms of the displacement of the gas element $\xi$. In additive, as it can be seen from the solution of the linearized problem (subsection 3.2), for sufficiently large $r$ the differential rotation law does not determine directly the dependence $\xi(t, r)$, but enters only the phase of the oscillating term. In reality, the role of differential rotation will be manifested in the fact that the distribution of the unperturbed density and the speed of sound are inhomogeneous in space and the law of this distribution is dictated, among other factors, by the law of differential rotation (see \ref{sec:equillibrium}).

We reduced the gas flow equations to the one-dimensional Euler equation and continuity equation. It is shown in \cite{Landau1987theorphys-6} that an arbitrary one-dimensional isentropic gas flow with the adiabatic index $\gamma$ may be described by a linear equation of the form
\begin{equation}
  \label{eq:chi}
  (\gamma-1)\,w\,\DPfrac{^2\chi}{w^2}
    - \DPfrac{^2\chi}{v^2}
    + \DPfrac{\chi}{w}
  = 0  \;,
\end{equation}
where $v$ is the velocity; $w = \int dp/\rho = c^2/(\gamma-1)$ --- enthalpy. Solution of this equation, the function $\chi(v, w)$, is defined as
\begin{equation}
  \chi
  = \varphi - v r + \left( \frac{v^2}{2} + w \right) t  \;,
\end{equation}
where $\varphi$ is velocity potential, i.e. $v = \partial \varphi/\partial r$. Time $t$ and coordinate $r$ are expressed as:
\begin{gather}
  \label{eq:chi_t_general}
  t
  = \DPfrac{\chi}{w}  \;,  \\
  \label{eq:chi_r_general}
  r - v t
  = - \DPfrac{\chi}{v}  \;.
\end{gather}

We are not interested in the complete solution of the problem of nonlinear evolution of a perturbations with a given initial form, but only in the analysis of the possibility of formation of discontinuities in it. The location of the discontinuity can be defined as the point at which the plots of the velocity and/or enthalpy dependence on the coordinate become multivalued. In terms of the function $r(v, w)$ the position of discontinuity corresponds to the inflection point:
\begin{equation}
  \label{eq:shock_conditions_v}
  \TDfrac{r}{v}{t} = 0  \;,\qquad
  \TDfrac{^2r}{v^2}{t} = 0  \;,
\end{equation}
while the derivatives are computed at the fixed instant of the time. It is easy to see that the r.h.s. of (\ref{eq:chi_r_general}) describes dependence of $r(v, w)$ at $t = 0$; we will denote it as  $R(v, w)$. The first of expressions (\ref{eq:shock_conditions_v}) together with (\ref{eq:chi_r_general}) provides the time of formation of discontinuity in solution:
\begin{equation}
  t_\mathrm{sh}
  = - \TDfrac{R}{v}{t}  \;.
\end{equation}
The calculation of this derivative in the general case may turn out to be a non-trivial task. However, we can express the position of the element we can express position of a gas element $R$ кas a function of velocity, by inversion of the dependence of the velocity on coordinates at $t = 0$. Let take as an initial perturbation configuration a section of the profile of a linear wave (\ref{eq:v_approx}):
\begin{equation}
  v|_{t=0}
  = - \mathcal{M}_\ast c_\ast\,\frac{r}{r_\ast}
      \,\sin\!\left[ \frac{2 r_\ast \omega}{3 c_\ast} \left( \frac{r}{r_\ast} \right)^{3/2} \right]  \;,
\end{equation}
where $\mathcal{M}_\ast$ is Mach number at $r_\ast$. Without limiting generality, let select $r_\ast$ as a point, where velocity becomes zero and declines in the direction of progression of the wave. It is evident that this point is also inflection point for dependence $r(v)$. Let consider the wave in a small vicinity of this point. Let introduce $x = r - r_\ast$; it will not be an additiveal restriction to require satisfaction of condition $|x| \ll r_\ast$:
\begin{equation}
  v|_{t=0}
  = - \mathcal{M}_\ast \omega x
    + \mathcal{O}\!\left( \frac{x^2}{r_\ast^2} \right)  \;.
\end{equation}
Now, with precision down to small quantities of the second order, we may write down:
\begin{equation}
  R
  = r_\ast - \frac{v}{\mathcal{M}_\ast \omega}  \;.
\end{equation}
Inserting this expression into (\ref{eq:chi_r_general}) and applying the first of conditions (\ref{eq:shock_conditions_v}) (the second one is fulfilled automatically), we obtain an expression for the time of discontinuity formation:
\begin{equation}
  \label{eq:shock_time}
  t_\mathrm{sh}
  = \frac{1}{\mathcal{M}_\ast \omega}  \;,
\end{equation}
where $\mathcal{M}_\ast$ and $\omega$ should be interpreted as parameters of perturbation in the point $r_\ast$.

It is easy to find the point of formation of discontinuity $r_\mathrm{sh}$, if we recall that the local phase velocity of small-scale perturbations is equal to the sound velocity. In the general case of an arbitrary direction of progression of the wave we obtain
\begin{equation}
  \label{eq:shock_position_general}
  t_\mathrm{sh}
  = \pm \int_{r_\ast}^{r_\mathrm{sh}} \frac{dr}{c_0}  \;.
\end{equation}
The sign in the front of r.h.s. depends on direction of the wave propagation --- the outer edge of the disk (\glqq{$+$}\grqq)  or to the inner one  (\glqq{$-$}\grqq). Inserting (\ref{eq:shock_time}) and expression for the sound velocity $c_0 = c_\ast\,(r/r_\ast)^{-1/2}$ into (\ref{eq:shock_position_general}), we obtain
\begin{equation}
  \label{eq:shock_position}
  r_\mathrm{sh}
  = r_\ast \left( 1 \pm \frac{3 c_\ast}{2 \mathcal{M}_\ast r_\ast \omega} \right)^{2/3}  \;.
\end{equation}

Note, for the waves progressing to the center of the accretion disk, the formula (\ref{eq:shock_position}) is valid in the wave region, i.e., as long as condition $\omega > \Omega_\mathrm{K}(r_\mathrm{sh})$ is valid (see subsection \ref{sec:linear_pert}). The time it takes for perturbation to reach the border of the wave zone may be estimated from the latter condition and expression (\ref{eq:shock_position_general}):
\begin{equation}
  t_\mathrm{sh,max}
  = \frac{2 r_\ast}{3 c_\ast} \left( 1 - \frac{\Omega_{\mathrm{K}\ast}}{\omega} \right)  \;.
\end{equation}

If the discontinuity did not form during this time, then, following energy conservation law, the wave should reflect from the wave region boundary and head to the outer part of the disk. In this case, the wave has one more chance to form the discontinuity. However, the very reflection of the wave from the boundary $\ell = H$ can not be described in terms of the present model.

It is evident that all above mentioned is valid also for dependence of perturbation of enthalpy on radius. At this, from the form of functions (\ref{eq:rho_approx}) and (\ref{eq:v_approx})  it is clear that inflection points for $r(v)$ and $r(w)$ coincide.

\begin{figure*}[ht!]
  \centering
  \includegraphics{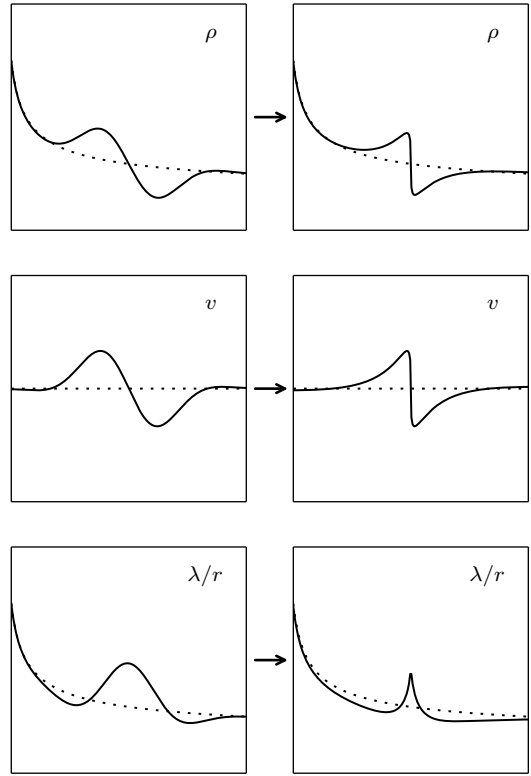}
  \caption{The diagram of formation of a discontinuity in a rotating flow. The perturbation propagates from left to right. \textit{Upper row:} the profile of density. \textit{Middle row:} the profile of radial velocity. \textit{Bottom row:} the profile of tangential velocity. Dotted line show equilibrium distributions.}
  \label{fig:shock_form}
\end{figure*}

It is important to note that the angular momentum distribution in the perturb flow will behave differently. Indeed, expressions (\ref{eq:rho_approx}), (\ref{eq:v_approx}) and (\ref{eq:lambda_approx}) show that, if the dependence of density and radial velocity passes through zero (the front of the perturbation wave), the angular momentum passes through the extremum. This means that in the plot of the dependence of the angular momentum on the radius does not appear discontinuity. However, since the angular momentum is transferred by the gas as a passive scalar, see (\ref{eq:angular_momentum_general}), at the instant of formation of the discontinuity of density and radial velocity, the distribution of angular momentum will get the shape of a caustic with a peak located at the point  $r_\mathrm{sh}$. Schematically this is shown in Fig. \ref{fig:shock_form}. As it is seen, the amplitude of the density and radial velocity step is determined by the change of the corresponding value over the scale of half wavelength, $\ell/2$, of the initial perturbation. The change of tangential velocity on the caustic is determined mainly by the difference of the values of the tangential velocity in the original distribution over the scale $\ell/4$.

\section{Application to accretion disks}

In the previous Section we obtained the estimates of the formation time of discontinuities in a rotating flow. As was shown in \cite{Landau1987theorphys-6}, discontinuities of density and velocity, formed as a result of rollover of the wave profile are asymptotically damped because of microscopic viscosity. However, in the case of a rotating flow, the picture of the evolution of a perturbation after formation of a discontinuity may change qualitatively.

We noted above that at the time of the formation of the discontinuous solution the tangential velocity profile takes the form of a caustic. In the neighborhood of the caustic tangential velocity will experience a strong change. Let $\delta \lambda$ be the difference of the angular moment in the initial tangential velocity distribution over the scale $\ell/4$:
\begin{equation}
  \delta \lambda
  \approx \left| \Dfrac{\lambda_0^2}{r}\,\frac{\ell/4}{2 \lambda_0} \right|_{r_\mathrm{sh}}
  \approx \frac{\Omega_\mathrm{K}(r_\mathrm{sh})}{8 \omega}\,r_\mathrm{sh} c_0(r_\mathrm{sh})  \;.
\end{equation}
Here, we neglected a small deviation from Keplerian distribution of angular momentum in the expression (\ref{eq:subkeplerian_momentum}). The change of tangential velocity at the caustic is equal to $\delta \lambda/r_\mathrm{sh}$ or, in the units of sound velocity,
\begin{equation}
  \label{eq:alpha}
  \alpha
  \equiv \frac{\Omega_\mathrm{K}(r_\mathrm{sh})}{8 \omega}
  = \frac{\Omega_{\mathrm{K}\ast}}{8 \omega} \left( 1 \pm \frac{3 c_\ast}{2 \mathcal{M}_\ast r_\ast \omega} \right)^{-1}  \;.
\end{equation}

In the reference frame connected with the surface of discontinuity, configuration of the flow resembles a laminar boundary layer of the \glqq{planar jet}\grqq. Indeed, the typical width of the caustic is much smaller than the wavelength $\ell$, which in the wave zone can not exceed semi-thickness of the disk. At the same time, in the tangential direction the flow is fairly homogeneous. It is well known that the flow of this type becomes unstable if the Reynolds number $\mathsf{Re} \approx 3.7$ (see, for example, \cite{Monin1986SvPhU..29..843M}). With increase of the Reynolds number, the range of the wave numbers of perturbations which cause the instability of the boundary layer expands without limits. For a typical accretion disk, one can estimate that $\mathsf{Re} \sim 10^{10}$ \cite{Razdoburdin2015PhyU...58.1031R}. Even if we determine the Reynolds number from the value of the tangential velocity difference (\ref{eq:alpha}), its estimate decreases by two or three orders of magnitude only. This allows to claim that the region of the caustic in the distribution of the tangential velocity is unstable with respect to perturbations of practically any scale. We note that the region of the caustics is located at both sides of the density and radial velocity jump. Since the jump is a shock wave, it propagates in the gas with supersonic velocity. By virtue of this, small-scale perturbations can propagate behind the shock only \cite{Landau1987theorphys-6}, and, therefore, the instability can develop only there. Given that accretion disks are typically have huge Reynolds numbers, instability of the distribution of tangential velocity in the caustic may well become a source of turbulence in the disk. Then the quantity $\alpha$ (\ref{eq:alpha}) should be interpreted as an estimate of the Shakura-Sunyaev coefficient \cite{Shakura1973A&A....24..337S}.

\begin{figure*}[ht!]
  \centering
  \includegraphics{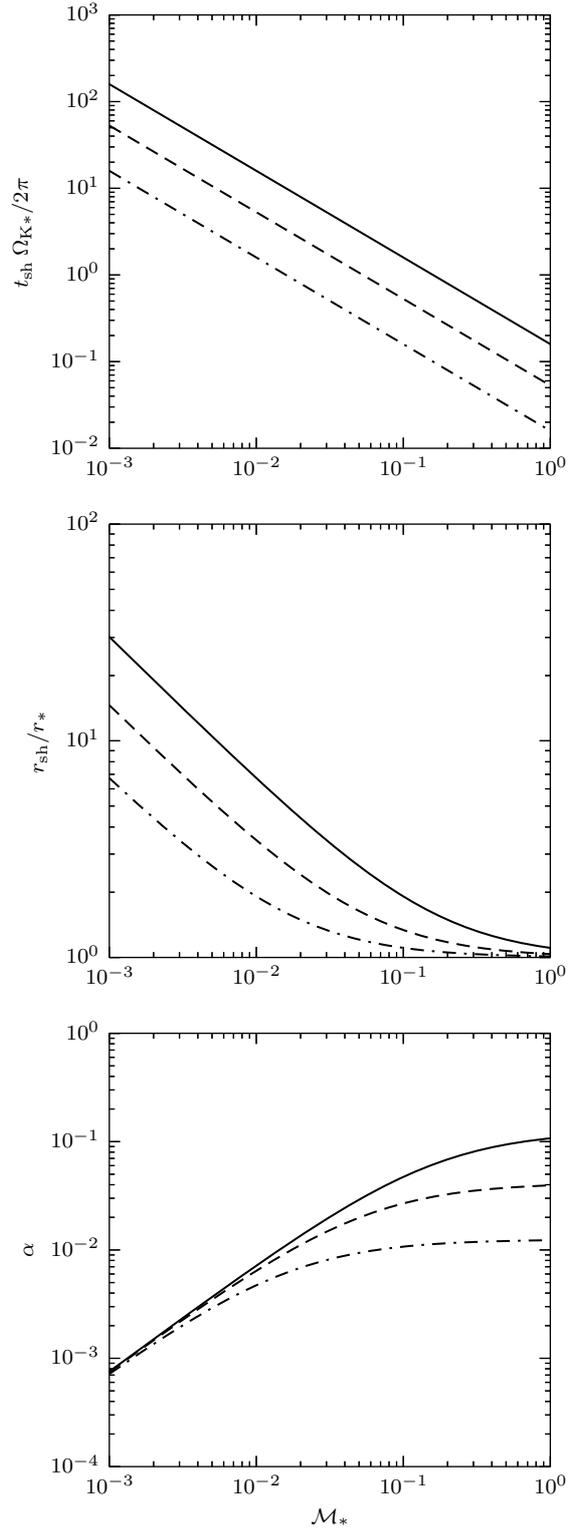}
  \caption{Dependence of the time, radius of formation, and amplitude of tangential velocity on the spatial scale and Mach number of the initial perturbation. Perturbation wave moves from the inner boundary of the disk ($r_\ast$) to the outer one. \textit{Solid line:} $\ell(r_\ast) = H(r_\ast)$. \textit{Dashed line:} $\ell(r_\ast) = H(r_\ast)/3$. \textit{Dash-dotted line:} $\ell(r_\ast) = H(r_\ast)/10$.}
  \label{fig:shock_out}
\end{figure*}

Expression (\ref{eq:alpha}) is written down for the waves that move from the center (sign \glqq{$+$}\grqq) and to the center (sign \glqq{$-$}\grqq). If the perturbation propagates toward the outer region of the disk, the value of $\alpha$ increases with increasing amplitude of the initial perturbation $\mathcal{M}_\ast$, but it can not exceed $1/8$ in principle due to condition (\ref{eq:wave_constraint}). Let us consider as an example a protoplanetary disk around binary system with the total mass $M_\mathrm{tot} = 2 M_\odot$, orbital separation $A = 10 R_\odot$ and the inner disk radius $r_\mathrm{in} = 2 A$. We set the temperature of the disk equal to $10^4$~К.  Let assume that the wave moves from the inner boundary of the disk to the outer one (in this case it is necessary to assume $r_\ast = r_\mathrm{in}$). Figure \ref{fig:shock_out} shows the plots of the dependence on time of the radius of formation of the discontinuity and the amplitude of the jump of tangential velocity on the spatial scale and Mach number of the initial perturbation. It is seen that for $\mathcal{M}_\ast \lesssim 10^{-2}$ the value of $\alpha$ weakly depends on the initial wavelength of perturbation, but increases with the amplitude as $\alpha \sim \mathcal{M}_\ast$. The radius, at which the discontinuity is formed and the time of its formation rapidly grow with increasing wavelength, i.e., long-wave perturbations of small amplitude can reach the outer boundary of the disk, without having had time to form a discontinuity. Perturbations of larger amplitudes, naturally, quite quickly form high intensity jumps, where $\alpha \lesssim 1/8$.

\begin{figure*}[ht!]
  \centering
  \includegraphics{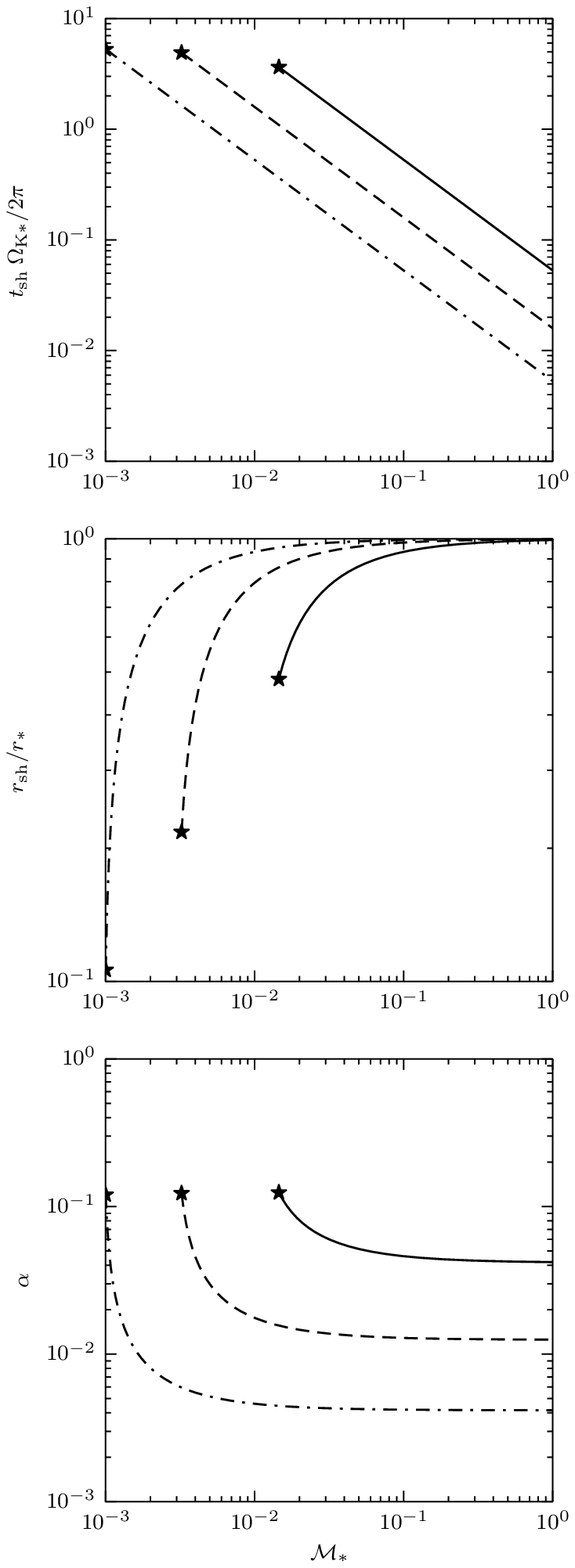}
  \caption{Dependence of the time, radius of formation, and amplitude of tangential velocity on the spatial scale and Mach number of the initial perturbation. Perturbation wave moves from the inner boundary of the disk ($r_\ast$) to the outer one. \textit{Solid line:} $\ell(r_\ast) = H(r_\ast)/3$. \textit{Dashed line:} $\ell(r_\ast) = H(r_\ast)/10$. \textit{Dash-dotted line:} $\ell(r_\ast) = H(r_\ast)/30$. By asterisks are shown the cases when the wave reaches the border of the wave zone.}
  \label{fig:shock_in}
\end{figure*}

Scenario of formation of discontinuities in the case when the wave comes from the outer side of the disk, is somewhat different. Let consider as an example a circumstellar disk around a white dwarf with the mass $M = 1 M_\odot$. Let set the inner radius of the disk equal to $r_\mathrm{in} = 0.03 R_\odot$ and the outer one to $r_\mathrm{out} = 0.3 R_\odot$, the temperature of the disk to $10^4$~К; $r_\ast = r_\mathrm{out}$. The parameters of the discontinuities that form in such a disk are shown in Fig. \ref{fig:shock_in}. Unlike the first case, here the time of propagation of the wave is limited by the time it intersects the wave zone (a situations when the discontinuity forms strictly at the boundary of the wave zone are marked in Fig. \ref{fig:shock_in} by asterisks). As the wave propagates into the disc, sound velocity increases. This explains a large gradient of the dependences $r_\mathrm{sh}(\mathcal{M}_\ast)$ and $\alpha(\mathcal{M}_\ast)$  in the neighborhood of the boundary of the wave zone. The value of $\alpha$ at this point attains its maximum value $1/8$. Perturbations of a small initial amplitude, as a rule, lead to higher values of the coefficient $\alpha$ than in the first case, up to two orders of magnitude. Wave region of small-scale perturbations can cover entire disk. If the amplitude of such perturbations is small, they can reach the inner boundary of the disk, before they form a discontinuity.

\section{Conclusion}

In this paper we studied the evolution of nonlinear perturbations in a single class of sub- Keplerian accretion disks. It was assumed that disks can be exposed to external action at inner and outer borders. The first case corresponds to protoplanetary disks around binary systems of young stars of T Tau type; the source of impact can be detached shock waves. The second case can be realized in accretion disks of close binary stars, where perturbations can be generated by the region of interaction of the accretion flow and the disk.

It is found that at the front of a radial longitudinal perturbation wave a discontinuity of the density and radial velocity forms with time, which is a shock wave. Distribution of the tangential velocity in this case takes the form of a caustic. Position of the peak ofthe latter corresponds to the position of the shock wave. Formation of the shock wave occurs due to the well-known mechanism of the \glqq{breaking}\grqq\ of the profile of the nonlinear perturbation. The value of the jump of the tangential velocity at the caustic depends on the amplitude and length of the perturbation wave. In some cases it can reach about $0.1$ of the sound velocity. Such a configuration is unstable and disintegrates. The natural consequence of the decay is turbulization of the accretion disk. The estimates of the Shakura-Syunyaev parameter for the accretion disks --- $\alpha < 0.1$ ---  in both considered cases indicate a substantially subsonic turbulence.

\section{Acknowledgments}

E.P. Kurbatov was supported by the Russian Foundation for Basic research (contracts No. 14-29-06059 and 15-02-06365).

\section{Appendix A}

After substitution $\eta = \rho_0 \xi$, Eq. (\ref{eq:xi}) for displacement of the gas elements becomes:
\begin{equation}
  \label{app:eq:eta}
  \DPfrac{^2 \eta}{t^2}
  = c_0^2\,\DPfrac{^2\eta}{r^2}
    + (\gamma - 2)\,\frac{c_0^2}{\rho_0}\,\DPfrac{\rho_0}{r}\,\DPfrac{\eta}{r}
    - \frac{1}{r^3}\,\Dfrac{\lambda_0^2}{r}\,\eta  \;.
\end{equation}
We do not provide boundary conditions for this equation, but seek solution in the form of propagating perturbance. Therefore, we may apply the standard method of separation of variables:
\begin{equation}
  \eta(t, r)
  = T(t)\,R(r)  \;.
\end{equation}
Then, denoting by $-\omega^2$ integration constant, we get (dot and prime mark denote differentiation over time and spatial coordinate, respectively)
\begin{equation}
  \frac{\ddot{T}}{T}
  = \frac{c_0^2}{R} \left[ R'' + (\gamma - 2)\,(\ln \rho_0)' R' \right]
    - \frac{1}{r^3}\,\Dfrac{\lambda_0^2}{r}
  = - \omega^2  \;.
\end{equation}
Let write down solution for the temporal part as
\begin{equation}
  T
  \propto e^{\mp i \omega t}  \;.
\end{equation}
For the spatial part we get an equation
\begin{equation}
  R''
    + (\gamma - 2)\,(\ln \rho_0)' R'
    + \left( \frac{\omega^2}{c_0^2} - \frac{1}{c_0^2 r^3}\,\Dfrac{\lambda_0^2}{r} \right) R
  = 0  \;.
\end{equation}
Let insert to the last Eq. radial profiles of the density and sound velocity obtained in subsection \ref{sec:equillibrium}. Then, after substitution of the variables, $x = r/r_\ast$, $R(r) = X(x)$, and introduction of denotations:
\begin{gather}
  \mu = \frac{r_\ast \omega}{c_\ast}  \;,  \\
  \varkappa = \frac{r_\ast \Omega_{\mathrm{K}\ast}}{c_\ast}  \;,
\end{gather}
we obtain:
\begin{equation}
  x^2 X''
    + \frac{x}{2}\,X'
    + \left( \mu^2 x^3 - \varkappa^2 + \frac{3}{2} \right) X
  = 0  \;.
\end{equation}
Solution of this equation will be expressed via Hankel function of the first type:\footnote{%
Hankel functions are related to the Bessel functions of the first and second type: H $H_\nu = J_\nu + i Y_\nu$.} \cite[с. 169]{Zaitsev2003ode.book}:
\begin{equation}
  \label{app:eq:x}
  X
  = C r_\ast \rho_\ast x^{1/4} H_\nu\!\left( \frac{2 \mu}{3}\,x^{3/2} \right)  \;,
\end{equation}
where $C$ is a complex constant; the index of Hankel function is equal to
\begin{equation}
  \nu
  = \frac{1}{6}\,(16 \varkappa^2 - 23)^{1/2}  \;.
\end{equation}

For larger arguments of Hankel function, $\mu x^{3/2} \gg 1$, one has $H_\nu(z) \approx (\pi z/2)^{-1/2} \operatorname{exp}[i z + i (\dots)]$, and then
\begin{equation}
  \label{app:eq:x_aprx}
  X
  \approx C r_\ast \rho_\ast \left( \frac{3}{\pi \mu} \right)^{1/2} x^{-1/2}
    \operatorname{exp}\left[ i \frac{2 \mu}{3}\,x^{3/2} + i (\dots) \right]  \;,
\end{equation}
where $i (\dots)$ --- insignificant phase addition.

\begin{figure*}[ht!]
  \centering
  \includegraphics{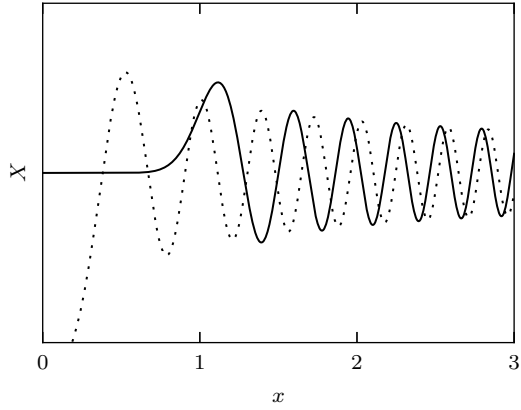}
  \caption{Typicalform of solution of the equation for radial displacemen of a Lagrangian particle. \textit{Solid line:} exact solution (\ref{app:eq:x}). \textit{Dotted line:} approximate solution (\ref{app:eq:x_aprx}). In point $x = 1$ condition (\ref{app:eq:wave_constraint}) is violated.}
  \label{fig:wave_cart}
\end{figure*}

Note, if solution has to have wave character, it is necesarry to satisfy the following inequality:
\begin{equation}
  \label{app:eq:wave_constraint}
  \mu^2 x^3 - \varkappa^2 + \frac{3}{2} > 0  \;.
\end{equation}
or
\begin{equation}
  x
  > \left( \frac{\varkappa^2}{\mu^2} - \frac{3}{2\mu^2} \right)^{1/3}  \;.
\end{equation}
For low $x$ solution (\ref{app:eq:x})  very rapidly aperiodically approaches zero. Figure \ref{fig:wave_cart}  shows approximately the dependence  $X(x)$.

Exact solution for linear perturbation may be obtained also for full set up of the problem, without neglect of the \glqq{geometrical}\grqq\ term in the continuity equation (\ref{eq:continuity_general}). In this case, the value of displacement and density will be related as
\begin{equation}
  \rho_1
  = - \frac{1}{r}\,\DPfrac{(r \rho_0 \xi)}{r}  \;.
\end{equation}
After substitution $\eta = r \rho_0 \xi$ we obtain an equation analogous to (\ref{app:eq:eta}):
\begin{equation}
  \DPfrac{^2 \eta}{t^2}
  = c_0^2 r\,\DPfrac{}{r} \left( \frac{1}{r}\,\DPfrac{\eta}{r} \right)
    + (\gamma - 2)\,\frac{c_0^2}{\rho_0}\,\DPfrac{\rho_0}{r}\,\DPfrac{\eta}{r}
    - \frac{1}{r^3}\,\Dfrac{\lambda_0^2}{r}\,\eta  \;.
\end{equation}
From this equation, by means of separation of variables and substitutions listed above, it is possible to get an equation for $X$, which differs from the last Eq. by the sign in the front of the second term only:
\begin{equation}
  x^2 X''
    - \frac{x}{2}\,X'
    + \left( \mu^2 x^3 - \varkappa^2 + \frac{3}{2} \right) X
  = 0  \;.
\end{equation}
Its solution \cite{Zaitsev2003ode.book}:
\begin{equation}
  X
  = C r_\ast^2 \rho_\ast x^{3/4} H_\nu\!\left( \frac{2 \mu}{3}\,x^{3/2} \right)  \;,
\end{equation}
where the index of hankel function is
\begin{equation}
  \nu
  = \frac{1}{6}\,(16 \varkappa^2 - 15)^{1/2}  \;.
\end{equation}
This solution does not differ essentially from Eq. (\ref{app:eq:x}), obtained in quasi-Cartesian approximation.

\clearpage

\bibliography{paper}
\bibliographystyle{unsrt}

\end{document}